# Dislocation-Free InGaN Nanoscale Light-Emitting Diode Pixels on Single Crystal GaN Substrate

Nirmal Anand[1], Sadat Tahmeed Azad[3], Christy Giji Jenson[1], Dipon Kumar Ghosh[1], Md Zunaid Baten[1], Pei-Cheng Ku[4], Grzegorz Muziol[2] and Sharif Sadaf[1*]

[1]*Centre Energie, Matériaux et Télécommunications, Institut national de la recherche scientifique (INRS-EMT), Varennes, Québec J3X 1P7, Canada.*

[2]*Institute of High-Pressure Physics, Polish Academy of Sciences, Sokolowska 29/37, 01-142 Warsaw, Poland*

[3]*Department of Electrical and Electronic Engineering, Bangladesh University of Engineering and Technology (BUET), West Palashi, Dhaka 1205, Bangladesh.*

[4]*Department of Electrical Engineering and Computer Science, University of Michigan, Ann Arbor, Michigan 48109, United States*

[*]*: E-mail : sharif.sadaf@inrs.ca*

# TABLE OF CONTENT (TOC) GRAPHIC

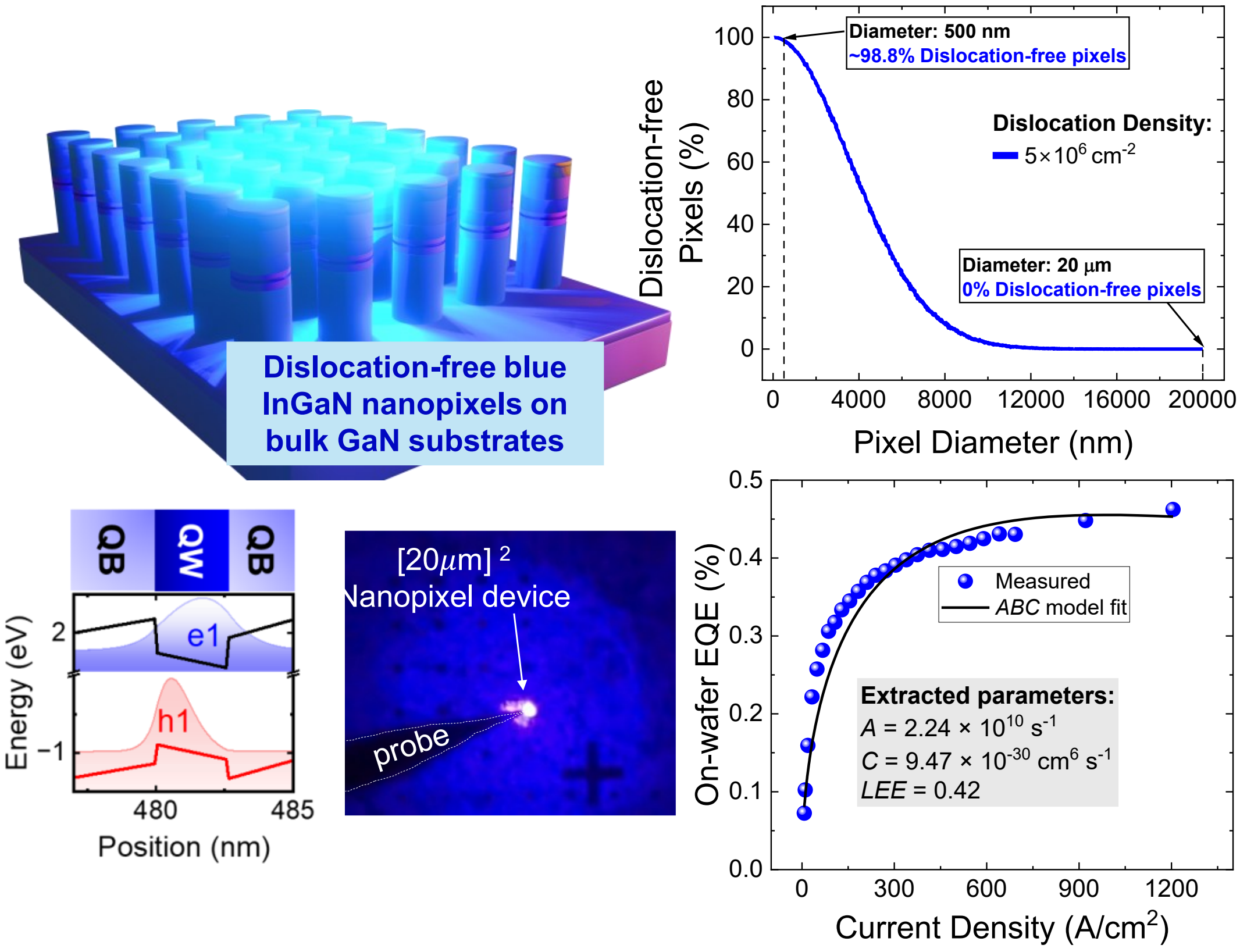

**ABSTRACT:**

Indium gallium nitride (InGaN) quantum well (QW) micro- and nanoscale light-emitting diodes (LEDs) are promising for next-generation ultrafast optical interconnects and augmented/virtual reality displays. However, scaling to nanoscale dimensions presents significant challenges including enhanced nonradiative surface recombination, defect and/or dislocation related emission degradation and nanoscale pixel contact formation. In this work, we demonstrate strain-engineered nanoscale blue LED pixels fabricated via top-down nanostructuring of an all-InGaN quantum well/barrier heterostructure grown by plasma-assisted molecular beam epitaxy (PAMBE) on significantly low dislocation-density single-crystal GaN substrates (in the order of ~$10^5$–$10^6$ $cm^{-2}$; 2 to 3 orders of magnitude lower than commercial GaN/Sapphire templates). Sidewall passivation using atomic layer deposition (ALD) of $Al_2O_3$ enables excellent diode behavior, including a rectification ratio >$10^4$ between –5 V and +5 V and extremely low reverse leakage (~0.01 A/$cm^2$ at –10 V). Monte Carlo analyses suggest almost 100% yield of completely dislocation-free active regions for ~450 nm nanopixels. Electroluminescence measurements show bright blue emission with a peak external quantum efficiency (EQE) of 0.46% at ~1.2 kA/$cm^2$. Poisson–Schrödinger simulations reveal ~20% strain relaxation in the QW, effectively mitigating the quantum-confined Stark effect (QCSE). Additionally, finite-difference time-domain (FDTD) simulations confirm that the nanoscale geometry enhances light extraction efficiency by over 40% compared to planar designs, independent of substrate materials. These results establish a scalable pathway for dislocation-free, high-brightness InGaN $\mu$LED arrays suitable for advanced display and photonic systems.

## 1. INTRODUCTION

Indium gallium nitride (InGaN) quantum well (QW) based light sources are crucial in advancing modern optoelectronics, particularly in ultrahigh-speed optical interconnects[1, 2], and cutting-edge near-eye displays for augmented and virtual reality (AR/VR)[3-5]. Such applications demand high pixel densities featuring micro- and nanoscale pixel dimensions, fast response times with multi-GHz modulation bandwidth, and superior brightness exceeding $10^3$ cd/$m^2$ [6-8]. Moreover, substantial efforts are underway to develop full-color (RGB) micro-displays with InGaN QW emitters. Despite the critical challenges inherent to micro- and nanoscale LEDs[9, 10], where an increased surface-area-to-volume ratio elevates the nonradiative Shockley–Read–Hall (SRH) recombination at the sidewalls, Samsung has recently achieved a record breakthrough by demonstrating a blue LED with nanoscale dimensions with a ~20.2% peak external quantum efficiency (EQE) through sol-gel based dielectric sidewall passivation[5].

However, to date, achieving efficient long-wavelength emission (green-red) with InGaN QWs remains a formidable challenge due to the green gap phenomenon—a significant drop in internal quantum efficiency (IQE) in the green-red spectral range[11]. The green gap arises primarily from the high indium (In) content required for green and red emission, which introduces severe underlying material challenges. The 11% lattice mismatch between InN and GaN induces a strong compressive strain in the QW region, aggravating the quantum-confined Stark effect (QCSE), which spatially separates electron and hole wavefunctions, thereby reducing radiative recombination efficiency[12, 13]. Additionally, high In incorporation leads to increased defect formation, phase separation, and In-rich clusters, all of which further degrade the optoelectronic performance[14, 15]. The inherently high threading dislocation densities (TDDs) on the order of $10^8$ to$10^9$ $cm^{-2}$ in commercially available GaN templates on c-plane sapphire substrates, where the InGaN heterostructures are grown, worsen the situation[16, 17]. These dislocations act as nonradiative recombination centers (NRCs) and inhibit uniform In incorporation, making it difficult to achieve stable and efficient green and red emission. Furthermore, InGaN heterostructures grown on GaN templates on sapphire substrates often exhibit a high density of V-pits, which can influence carrier recombination dynamics and introduce further nonuniformities in the emission properties[18, 19].

To overcome the above-mentioned critical challenges, several alternative epitaxial strategies have been explored, aimed at mitigating the TDD and V-pit related defects and addressing the green gap. One promising approach involves bottom-up growth of three-dimensional nanostructures [20, 21], which, due to their high surface-to-volume ratio, facilitate effective strain relaxation at the sidewalls, thereby reducing QCSE and improving In incorporation[22-24]. However, such bottom-up grown nanowire heterostructures often suffer from nonuniform structural morphology and varying carrier recombination lifetimes, particularly due to inhomogeneous indium incorporation at the faceted top surfaces[20, 25]. Recently, Wang *et al.* showed that top-down nanostructuring of LEDs after growth in the planar form, results in 90% dislocation free LEDs while maintaining IQE comparable to their planar counterparts[26]. Furthermore, previous reports have also shown that nanostructuring can alleviate compressive strain in InGaN QWs and enhance radiative recombination efficiency[27, 28]; however, most of these reports have focused on InGaN QWs with GaN barriers grown via metal-organic chemical vapor deposition (MOCVD).

In this report, we introduce a unique hybrid approach that combines plasma-assisted molecular beam epitaxial growth on single-crystal GaN substrates with subsequent top-down nanostructuring in InGaN heterostructures. By applying this technique to PAMBE-grown LED heterostructures with both InGaN QWs and barriers, we address fundamental challenges such as TDD-induced leakage current paths and V-pit-driven indium nonuniformity that limit efficient long-wavelength emission and uniform device performance. We demonstrate that our nanoscale LEDs exhibit excellent diode rectification and light output characteristics, attributed to improved current transport, enhanced carrier injection and enhanced light extraction resulting from the nanopixel architecture. This enables high-power operation by sustaining maximum EQE at high injection levels, which is crucial for energy-efficient $\mu$LED applications in ultrahigh-speed optical interconnects and near-eye displays.

## 2. EXPERIMENTAL

In this study, the blue-emitting InGaN single QW with InGaN barrier LED structure was grown using PAMBE on c-plane (0001) Ga-polar *n*-GaN bulk crystals with an ultra-low TDD of ~$10^6$ $cm^{-2}$. The high-quality InGaN layers were grown using a high active nitrogen flux RF plasma source[29-31]. The full PAMBE-grown LED epitaxial structure is shown in Figure 1 (a) and consists of a 450 nm *n*-type $In_{0.01}Ga_{0.99}N$:Si layer, a 2.6 nm $In_{0.179}Ga_{0.821}N$ single QW layer embedded by 30 nm and 20 nm $In_{0.021}Ga_{0.979}N$ barrier layers. Next, a 20 nm *p*-type $Al_{0.138}Ga_{0.862}N$:Mg electron blocking layer was grown, followed by 200 nm thick *p*-type GaN:Mg, 40 nm *p*-type $In_{0.022}Ga_{0.978}N$:Mg layer and finally a 5 nm *p*-type $In_{0.149}Ga_{0.851}N:Mg^+$ contact layer. The energy band diagram of the PAMBE-grown LED structure under equilibrium conditions (i.e., at 0 V bias) is shown in Figure 1(d), illustrating the built-in electric field arising from spontaneous and piezoelectric polarization effects. The simulation was performed using the one-dimensional drift-diffusion model implemented in the SiLENSe software package[32], as will be discussed in detail in later sections.

To fabricate the nanoscale LED arrays, the planar LED structure was patterned using electron beam lithography (EBL) to create ultra-dense circular pixels with a ~450 nm diameter and an intrapixel pitch of 700 nm. These arrays were arranged in a [20×20] $\mu m^2$ square lattice configuration. Next, 80 nm nickel (Ni) was e-beam evaporated onto the sample and lifted-off to be used as the etch mask for the top-down fabrication process. Subsequently, a $Cl_2$-based inductively coupled reactive-ion etch (ICP-RIE) recipe with optimized $Cl_2$/Ar gas flow was employed for the nanoscale pixel formation. Figure 1 (c) shows the scanning electron micrographs (SEM) of the nanoscale pixel array post ICP-RIE etch. The sidewalls appear tapered and smooth; however, they are typically associated with a high density of plasma-induced defects[33, 34]. The nanoscale pixel arrays were then wet etched with a buffered potassium hydroxide (KOH) solution at 80°C in order to effectively remove the plasma-induced amorphous sidewalls. Figure 1 (d) shows the SEM images of the nanoscale pixel array after wet etching. To further minimize the sidewall related leakage currents[35] and effectively passivate the remaining dangling bonds[36], the nanoscale pixel arrays were conformally coated with a 60 nm atomic layer deposited (ALD) $Al_2O_3$. In order to planarize the nanoscale pixel array, we utilized spin-on glass (SOG) coating. The $Al_2O_3$ and SOG layers were then etched-back using a combination of dry and wet etch techniques. Specifically, a $CF_4$ based ICP-RIE recipe was used to etch away the SOG, and a buffered tetramethylammonium hydroxide (TMAH) solution to etch the ALD-deposited $Al_2O_3$ to reveal the *p*-type InGaN contact layer on the top of the nanopixel array. A high-magnification SEM image is shown in Figure 2 (a).To define the *n*-contact openings, a wet etch using buffered hydrofluoric acid (HF) was performed, selectively removing the SOG and $Al_2O_3$ layers. Finally, a standard photolithography process was used to define the $[20\times20\ \mu m]^2$ device regions, to parallelly address the nanopixel array. A metal stack of 20 nm/40 nm Ni/Au was then deposited via e-beam evaporation to form the *p*-contacts. A Ti/Au stack of 20 nm/ 20 nm was used as the *n*-Contact. Contact annealing was performed at 500 °C for 5 min in a $N_2$ ambient. A 70° tilt-view SEM image of InGaN nanopixel array device stack is shown in Figure 2 (b).

## 3. RESULTS AND DISCUSSION

First, the fabricated nanopixel devices were electrically characterized by measuring their current density–voltage (*J–V*) characteristics, as shown in Figure 2 (c). The representative *J–V* curve exhibits excellent diode rectification behavior. It is well known that the electrical characteristics of InGaN $\mu$LEDs are strongly influenced by sidewall treatments due to their high surface-to-volume ratio[35, 37]. The nanopixel device achieves a current density of 100 A/cm$^2$ at ~3.9 V, which is comparable to state-of-the-art blue InGaN $\mu$LEDs with similar device areas. Importantly, the device exceeds 30 A/cm$^2$ at biases below 5 V, making it highly suitable for near-eye display applications[38]. Moreover, the leakage currents are extremely low, at 0.01 A/cm$^2$ at –10 V biasing, as shown in the inset of Figure 2 (c). Such low leakage currents can be directly attributed to the low density of extended defects[16, 17] (threading dislocations, V-pit etc.), the high material quality of the InGaN QW[39, 40] and excellent ALD-deposited dielectric sidewall passivation[41]. It is worthwhile mentioning that the excellent diode characteristics with low turn voltage and leakage current suggest that the current path and/or transport mechanism in the nanowire heterostructure LEDs grown on bulk GaN substrate grown by PAMBE is different from MOCVD grown heterostructures. Moreover, our study suggests that the forward voltage and leakage paths are not governed by the presence of V-pits and dislocation related tails states, which is in direct contrast to the MOCVD grown samples[42]. The current transport and I-V characteristics in dislocation and V-pit free heterostructures grown by PAMBE on GaN substrate is rather governed by the composition, thickness and number of the QWs, while the I-V characteristics largely depends on the dislocation related tail states, random alloy fluctuations, V-pits and the QWs for the MOCVD grown samples on foreign substrates: sapphire or silicon[43]. To further analyze the transport mechanism, we extracted the diode ideality factor '*n*' from the forward-bias region of the current-voltage curve. The ideality factor provides insights into the dominant carrier transport processes. For our nanopixel device, we obtained n ~3.5 near the turn-on voltage. While ideality factors close to 1 are indicative of diffusion-limited transport and values near ~2 suggest recombination in the space-charge region, higher ideality factors (n >> 2) are commonly observed in InGaN-based LEDs[44]. In InGaN nanowire-based LEDs, high ideality factors often arise from junction imperfections, including deep-level-assisted tunnelling, moderate doping, and metal–semiconductor contact resistances (e.g., Ni/Au to p-GaN)[45]. The cumulative effects lead to a high ideality factor. However, the moderate *n*-value in our InGaN nanopixel device, coupled with low leakage current and high rectification ratio (~$10^4$ from -5 V to +5 V), actually indicates good junction quality, likely due to the dislocation-free active region and effective ALD sidewall passivation, contrasting with more defective MOCVD-grown heterostructures on foreign substrates such as Sapphire and Si.To further investigate the role of dislocation density in device performance, we employed a Monte Carlo simulation to statistically evaluate the probability of dislocation-free regions within the parallelly-addressed nanopixel architecture[26]. As shown in Figure 3 (a), for an initial dislocation density of ~$5\times10^6$ $cm^{-2}$, assumed densities in HVPE-grown bulk GaN substrates, the InGaN nanopixels with diameters of 450 nm exhibit a predicted dislocation-free yield of ~98.8%. This suggests that nanostructuring at this scale (i.e. around ~500 nm diameter) is highly effective in isolating threading dislocations, resulting in virtually dislocation-free active regions. Importantly, although electron beam lithography (EBL) was used for nanopatterning in this study, the ~500 nm feature sizes demonstrated here are well within the resolution capabilities of industry-compatible deep ultraviolet (DUV) lithography as well as nanoimprint lithography (NIL), making this approach scalable and suitable for high-volume manufacturing[5, 46]. Figure 3 (b) shows the Monte Carlo simulations extended across a range of dislocation densities ($5\times10^{6}$ to $1\times10^{9}$ $cm^{-2}$), showing the influence of initial epi-layer dislocation density on the minimum diameter needed for >90% dislocation-free pixels. Figure 3(c) schematically illustrates the dislocation-isolation mechanism enabled by top-down nanostructuring, comparing two cases: one starting from a high threading dislocation density LED epi, and the other from a low threading dislocation density LED epi. In both scenarios, through top-down nanofabrication, threading dislocations are statistically reduced and/or excluded from the active regions by spatial filtering/confinement.

The electroluminescence (EL) spectra of the nanopixel devices were measured as a function of various current injection densities. EL emission was collected normal to the device surface from the top with an optical fiber to a UV-visible spectrometer to record the spectra. Figure 4 (a) shows the recorded EL spectra of the nanopixel device, with bright blue spectrum. The peak wavelengths and the full-width at half maximum (FWHM) of the various spectra were extracted via Gaussian fit and plotted in Figure 4 (b). The FWHM of the nanopixel device varies from ~45 nm at low injection current densities, to ~25 nm at high injection current densities. We observe a ~14 nm peak wavelength blue-shift when the injection current increases two to three orders of magnitude (from ~30 $A/cm^2$ to 1200 $A/cm^2$). This key observation is in direct contrast with prior studies that nanostructuring-induced strain relaxation would stabilize emission wavelengths by mitigating QCSE[28, 47]. Our findings indicate a more complex interplay between the barrier composition (i.e. InGaN barriers vs. conventional GaN barriers) and residual compressive strain, which can subtly influence the internal piezoelectric fields.

Our recent scanning X-ray diffraction microscopy (SXDM) studies [48] reveal that PAMBE grown, top-down processed micro- and nanostructured InGaN QWs with InGaN barriers exhibit strain gradients, i.e. partially relaxed edges vs. strained centers. Furthermore, we showed that the degree of relaxation is with respect to the barrier material, i.e. InGaN barrier. Similarly, in this work, the InGaN QW is relaxed relative to the InGaN barriers but retains localized strain variations. These residual and spatially varying strain fields can modify the energy bandgap and alter carrier confinement potentials. In particular, spatial strain variations across the nanopixel (center-to-edge) may lead to inhomogeneities in emission energy, intensity, and efficiency; especially as carrier screening effects become prominent at elevated injection levels.

To further understand the impact of strain relaxation on the PAMBE-grown single QW LED structure, we performed Poisson–Schrödinger simulations using the SiLENSe software package to investigate changes in the energy band profile and wavefunction distribution[32]. The simulated structure replicates the same heterostructure depicted in Figure 1 (a). A threading dislocation density (TDD) of $10^6$ $cm^{-2}$ was assumed for all layers. In the simulations, the degree of relaxation (DOR) in the $In_{0.179}Ga_{0.821}N$ QW was varied from 0% (fully strained) to 50%, to model the effects of strain relaxation via top-down nanostructuring. For the partially relaxed cases (10%–50% DOR), the surrounding InGaN barriers were assumed to be fully relaxed (100% DOR), while, for the fully strained QW case (0% DOR), the barriers were also assumed to be fully strained to simulate a fully strained as-grown thin film structure. All simulations were carried out under equilibrium conditions (0 V bias) to isolate and quantify the effects of strain relaxation (e.g. via top-down nanostructuring) on the reduction of the QCSE, while deliberately excluding any carrier-induced screening effects. Figure 5 (a) shows the simulated energy band profile for various DOR (0–50%) for the $In_{0.179}Ga_{0.821}N$ QW at 0 V bias of the PAMBE-grown LED heterostructure in this work (as shown in the inset). The default material properties provided with SiLENSe had been used. As expected, the energy band slope across the QW decreases with increasing DOR, indicating a reduction in the internal piezoelectric field. This results in an increase in emission energy of the $In_{0.179}Ga_{0.821}N$ QW due to the suppression of the QCSE. The electron and hole wavefunction overlap corresponding to the ground bound state $< e1h1 >$ were also simulated. Figure 5 (b) shows the wavefunctions for different DOR values. In the fully strained case (0% DOR), the band profile forms a triangular potential well, leading to significant spatial separation between electron and hole wavefunctions. As the DOR increases, the potential profile becomes more symmetric and uniformly confined (rectangular) due to weakening piezoelectric fields, leading to improved wavefunction overlap. Figure 5 (c) shows the carrier wavefunction overlap of $< e1h1 >$ bound states (left axis) with respect to the various degrees of relaxation (DOR) of the $In_{0.179}Ga_{0.821}N$ QW, along with the EL peak wavelength at 1 A/cm$^2$ (right axis).

To correlate our theoretical findings to practical device behavior, we simulated the EL spectra at low (~1 A/cm$^2$) and high (~1000 A/cm$^2$) current injection levels for different DOR values, and the results are summarized in Table 1. The blue-shift decreased systematically with increasing relaxation; from 17.25 nm at 0% DOR to 11.05 nm at 50% DOR, with reduced QCSE. Our experimental nanopixel LEDs exhibited a ~14 nm blue-shift, closely matching the simulated result for 20% DOR. This suggests that the fabricated $In_{0.179}Ga_{0.821}N$ QW in the nanopixel $\mu$LED is about ~20% relaxed due to the strain relaxation induced by top-down nanostructuring. It is worth noting that that using our nanostructuring approach we can controllably relax the QWs by changing the diameter of the nanowires.

For optical output power measurements, the emitted photons from the nanopixel devices were collected through the backside of the bulk *n*-GaN substrate. The hydride vapor phase epitaxy (HVPE)-grown *n*-type GaN substrate has high transparency across the visible spectrum, allowing for efficient backside light extraction. A calibrated Thorlabs S120VC photodiode was used to accurately measure the optical power output. The measured light output power density is shown in Figure 6 (a). We measure a ten-fold (10×) increase in light output power for the nanopixel array devices in comparison to the planar $\mu$LED devices grown using the same heterostructure with PAMBE[49]. The order-of-magnitude higher optical output power observed in nanopixel array devices compared to planar $\mu$LEDs can be attributed to a combination of several factors. First, the enhanced light extraction resulting from the nanostructured geometry of the nanopixels. Second, there is improved current spreading within the *p*-type (In)GaN layers of the nanopixel array, which reduces current crowding[50] and lateral carrier diffusion[51]. And finally, the experimental configuration that enables efficient backside photon collection through the highly transparent HVPE-grown *n*-GaN substrate also contributes to the high measured optical output.

Such device-level and experimental improvements collectively contribute to overall increased radiative recombination rates and also significantly higher measured optical output from the nanopixel architecture. Subsequently, the on-wafer external quantum efficiency (EQE) was calculated from the measured light output power and plotted in Figure 6 (b). A peak EQE of 0.46% was recorded for the parallelly addressed 500-nm nanopixels at a high current density of 1200 A/cm$^2$. Interestingly, no evidence of EQE droop is observed even at these elevated injection levels, which is in direct contrast with MOCVD-grown devices where droop typically becomes prominent at relatively lower current densities. Such apparent absence of droop can be due to the extreme non-equilibrium growth of the heterostructure by PAMBE. Although the LED heterostructure was grown at a relatively high growth temperature with N-rich growth conditions, the growth regime remains far from thermodynamic equilibrium, which induces significant Shockley-Read-Hall (SRH) trap/defect states. This also suggests that, in addition to nonradiative surface recombination, there could be significant point defects present in PAMBE grown InGaN heterostructures as opposed to MOCVD growth.

To gain deeper insights, a detailed analysis was performed for the EQE of the nanopixel device. The conventional ABC model was used to fit the measured EQE to derive the actual values of *A*, *C* and the light extraction efficiency (LEE)[9]:

$$EQE = \frac{B \cdot n^2}{A \cdot n + B \cdot n^2 + C \cdot n^3} \times LEE \qquad (1)$$

where *A* is the nonradiative Shockley-Read-Hall (SRH) recombination coefficient, *B* is the radiative recombination coefficient, *C* is the nonradiative Auger-Meitner recombination coefficient and *n* is the carrier concentration. To fit the experimental EQE data, and to extract the *LEE*, *A* and *C* coefficients, we assumed a radiative recombination coefficient $B = 1 \times 10^{-11}$ cm$^3$ s$^{-1}$, which correspond to typical empirical values used for blue InGaN $\mu$LEDs. Consequently, from the fit, we extracted the coefficients *A* and *C* to be $2.24 \times 10^{10}$ s$^{-1}$ and $9.47 \times 10^{-30}$ cm$^6$ s$^{-1}$, respectively and estimate the LEE to be around 42% for the nanopixel array. The fitted curve shows good agreement with the experimental data and indicates that SRH recombination is the dominant non-radiative recombination mechanism[52].

While these results emphasize the effectiveness of our PAMBE-grown top-down nanopixel architecture in achieving high light output operation, further theoretical studies are required to comprehensively understand the underlying mechanisms for the light output/collection enhancement. These include Purcell enhancement, surface plasmon coupling, polarization dynamics, symmetry-induced emission properties, and the influence of substrate surface roughness[36, 53-56]. To this end, finite difference time domain (FDTD) simulations have been performed to understand the enhanced light extraction/collection characteristics of the fabricated nanopixel array, particularly the advantage of using GaN substrates over a more commercially viable sapphire substrate. The FDTD simulations were performed using Ansys Lumerical FDTD package[57]. These simulations were designed to investigate the impact of the nanostructure geometry (i.e., 450 nm diameter nanopixel array, in a square lattice with a 700 nm pitch to emulate the fabricated device) on the light extraction performance. A simulation area of 5 $\mu$m × 5 $\mu$m was defined, where InGaN nanopixels (refractive index ≈ 2.6) with heights of 1 $\mu$m and diameters of 450 nm were placed on top of a 1 $\mu$m thick GaN layer (refractive index ≈ 2.4). This structure was then modelled on top of either a GaN or a sapphire substrate to directly compare substrate-dependent light extraction behavior. A single dipole source (TE-mode) was placed at 200 nm from the top of each nanopixel. A schematic of the modelled simulation is shown in Figure 7 (a). A light collection monitor was placed beneath the substrate to record the transmitted optical power (i.e., photons) through the backside, replicating the experimental measurement configuration. The collected power was then normalized to the total emitted power from the dipole to calculate the extraction efficiency over the blue wavelengths (445 nm to 455 nm). As shown in Figure 7 (b), our simulated results show that the nanopixel architectures improve light extraction by ~40% over planar designs in the 445 nm to 455 nm range, regardless of substrate, within the blue spectral region. However, GaN substrates enable significantly higher quality InGaN growth due to their ultra-low dislocation densities for their use in high-performance $\mu$LED applications. We have also investigated the directionality of emission to provide deeper insight into the nanostructure-mediated light extraction mechanisms. We examined the far-field radiation pattern of the nanopixel array on GaN substrate, as shown in Fig. 7 (c). The simulation reconstructs the far-field distribution by extending the collected near-fields as plane waves projected onto a hemispherical surface with a ~1 m radius centered around the device. The radial angle ($\theta = 0°$–$90°$) is defined relative to the vertical z-axis, while the azimuthal angle ($\phi = 0°$–$360°$) is measured in the xy-plane, corresponding to the lateral (edge) direction of the device. The resulting radiation pattern reveals four prominent emission lobes located ~43° from the vertical axis along both x and y directions. These distinct lobes arise due to a combination of dipole source orientation and the periodic nanostructured geometry, which governs light redirection and diffraction. Our detailed simulation suggests that it is possible to control the light propagation direction, beam intensity and enhance light extraction efficiency by controlled nano-structuring of the LED heterostructure by top-down nanofabrication.

## 4. CONCLUSION

In summary, we have demonstrated dislocation-free InGaN strain-relaxed nanopixel LEDs fabricated through a top-down nanostructuring approach on bulk GaN substrates. The integration of low-dislocation-density substrates with atomic layer deposition (ALD) $Al_2O_3$ dielectric sidewall passivation enables superior material quality, resulting in ultra-low leakage currents and excellent diode rectifying behavior. The devices exhibit bright electroluminescence with a peak EQE of 0.46% at a high injection current density of ~1200 A/cm$^2$, which highlights their potential for high-brightness, high-current-density applications. Poisson–Schrödinger simulations suggest partial suppression of the quantum-confined Stark effect (QCSE) due to strain relaxation in the nanopixel QW geometry, while FDTD simulations reveal a ~40% enhancement in light extraction for the nanopixel array compared to planar geometries in the blue spectral region. These findings emphasize the effectiveness of nanoscale pixel engineering in overcoming the scaling challenges and paving the pathway for low-power $\mu$LED platforms suitable for ultra-high speed optical interconnects, next-generation displays, and emerging photonic technologies.

**ACKNOWLEDGEMENTS**

This work was supported by the Natural Sciences and Engineering Research Council of Canada (NSERC) through Discovery and Quantum Alliance Grant programs, MEIE Photonique Quantique Quebec (PQ2) program, and the Canada Research Chair program, and "Monolithic integration of multi-color arrays of micro- and nano-LEDs" project carried out within the First Team program of the Foundation for Polish Science co-financed by the European Union under the European Funds for Smart Economy 2021-2027 (FENG), and US National Science Foundation Award 2317047.

**REFERENCES**

(1) Ahmed, K. Techno-economics of axial nanowire light-emitting diodes for augmented reality displays and data communication applications. In Light-Emitting Devices, Materials, and Applications XXIX, 2025.

(2) Ahmed, K. Scaling theory of enhanced speed in N-Polar GaN MicroLEDs. In Physics and Simulation of Optoelectronic Devices XXXIII, 2025.

(3) Anand, N.; Ghosh, D. K.; Abbes, A.; Kundu, M.; Rahman, M. A.; Jenson, C. G.; Morandotti, R.; Baten, M. Z.; Sadaf, S. M. Ultra-dense Green InGaN/GaN Nanoscale Pixels with High Luminescence Stability and Uniformity for Near-Eye Displays. *ACS Nano* **2024**, *18* (39), 26882-26890.

(4) Chung, K.; Sui, J.; Demory, B.; Teng, C.-H.; Ku, P.-C. Monolithic integration of individually addressable light-emitting diode color pixels. *Applied Physics Letters* **2017**, *110* (11).

(5) Sheen, M.; Ko, Y.; Kim, D. U.; Kim, J.; Byun, J. H.; Choi, Y.; Ha, J.; Yeon, K. Y.; Kim, D.; Jung, J.; et al. Highly efficient blue InGaN nanoscale light-emitting diodes. *Nature* **2022**, *608* (7921), 56-61.

(6) Nami, M.; Rashidi, A.; Monavarian, M.; Mishkat-Ul-Masabih, S.; Rishinaramangalam, A. K.; Brueck, S. R. J.; Feezell, D. Electrically Injected GHz-Class GaN/InGaN Core–Shell Nanowire-Based μLEDs: Carrier Dynamics and Nanoscale Homogeneity. *ACS Photonics* **2019**, *6* (7), 1618-1625.

(7) Behrman, K.; Kymissis, I. Micro light-emitting diodes. *Nature Electronics* **2022**, *5* (9), 564-573.

(8) Park, J.; Choi, J. H.; Kong, K.; Han, J. H.; Park, J. H.; Kim, N.; Lee, E.; Kim, D.; Kim, J.; Chung, D.; et al. Electrically driven mid-submicrometre pixelation of InGaN micro-light-emitting diode displays for augmented-reality glasses. *Nature Photonics* **2021**, *15* (6), 449-455.

(9) Olivier, F.; Daami, A.; Licitra, C.; Templier, F. Shockley-Read-Hall and Auger non-radiative recombination in GaN based LEDs: A size effect study. *Applied Physics Letters* **2017**, *111* (2).

(10) Smith, J. M.; Ley, R.; Wong, M. S.; Baek, Y. H.; Kang, J. H.; Kim, C. H.; Gordon, M. J.; Nakamura, S.; Speck, J. S.; DenBaars, S. P. Comparison of size-dependent characteristics of blue and green InGaN microLEDs down to 1μm in diameter. *Applied Physics Letters* **2020**, *116* (7).

(11) Auf der Maur, M.; Pecchia, A.; Penazzi, G.; Rodrigues, W.; Di Carlo, A. Efficiency Drop in Green InGaN/GaN Light Emitting Diodes: The Role of Random Alloy Fluctuations. *Phys Rev Lett* **2016**, *116* (2), 027401.

(12) Li, X.; DeJong, E.; Armitage, R.; Armstrong, A. M.; Feezell, D. Influence of trap-assisted and intrinsic Auger–Meitner recombination on efficiency droop in green InGaN/GaN LEDs. *Applied Physics Letters* **2023**, *123* (11).

(13) Ley, R.; Chan, L.; Shapturenka, P.; Wong, M.; DenBaars, S.; Gordon, M. Strain relaxation of InGaN/GaN multi-quantum well light emitters via nanopatterning. *Opt Express* **2019**, *27* (21), 30081-30089.

(14) Hammersley, S.; Kappers, M. J.; Massabuau, F. C. P.; Sahonta, S. L.; Dawson, P.; Oliver, R. A.; Humphreys, C. J. Effects of quantum well growth temperature on the recombination efficiency of InGaN/GaN multiple quantum wells that emit in the green and blue spectral regions. *Applied Physics Letters* **2015**, *107* (13).

(15) Ho, I. h.; Stringfellow, G. B. Solid phase immiscibility in GaInN. *Applied Physics Letters* **1996**, *69* (18), 2701-2703.

(16) Wang, G.; Huang, J.; Wang, Y.; Tao, T.; Zhu, X.; Wang, Z.; Li, K.; Wang, Y.; Su, X.; Wang, J.; et al. Growth and characterization of micro-LED based on GaN substrate. *Opt Express* **2024**, *32* (18), 31463-31472.

(17) Chang, H.-M.; Lim, N.; Rienzi, V.; Gordon, M. J.; DenBaars, S. P.; Nakamura, S. Enhanced optical gain of c-plane InGaN laser diodes via a strain relaxed template with reduced threading dislocation density. *Optics Express* **2024**, *32* (20).

(18) Quevedo, A.; Wu, F.; Tsai, T.-Y.; Ewing, J. J.; Tak, T.; Gandrothula, S.; Gee, S.; Li, X.; Nakamura, S.; DenBaars, S. P.; et al. Dislocation half-loop control for optimal V-defect density in GaN-based light emitting diodes. *Applied Physics Letters* **2024**, *125* (4).

(19) Li, X.; DeJong, E.; Armitage, R.; Feezell, D. Multiple-carrier-lifetime model for carrier dynamics in InGaN/GaN LEDs with a non-uniform carrier distribution. *Journal of Applied Physics* **2024**, *135* (3).

(20) Concordel, A.; Rochat, N.; Quach, A. M. N.; Rouviere, J. L.; Jacopin, G.; Napierala, J.; Daudin, B. Inhomogeneous spatial distribution of non radiative recombination centers in GaN/InGaN nanowire heterostructures studied by cathodoluminescence. *Nanotechnology* **2023**, *34* (49).

(21) Song, K. M.; Kim, D. H.; Kim, J. M.; Cho, C. Y.; Choi, J.; Kim, K.; Park, J.; Kim, H. White light emission of monolithic InGaN/GaN grown on morphology-controlled, nanostructured GaN templates. *Nanotechnology* **2017**, *28* (22), 225703.

(22) Limbach, F.; Hauswald, C.; Lahnemann, J.; Wolz, M.; Brandt, O.; Trampert, A.; Hanke, M.; Jahn, U.; Calarco, R.; Geelhaar, L.; et al. Current path in light emitting diodes based on nanowire ensembles. *Nanotechnology* **2012**, *23* (46), 465301.

(23) Sadaf, S. M.; Ra, Y. H.; Nguyen, H. P.; Djavid, M.; Mi, Z. Alternating-Current InGaN/GaN Tunnel Junction Nanowire White-Light Emitting Diodes. *Nano Lett* **2015**, *15* (10), 6696-6701.

(24) Sadaf, S. M.; Ra, Y. H.; Szkopek, T.; Mi, Z. Monolithically Integrated Metal/Semiconductor Tunnel Junction Nanowire Light-Emitting Diodes. *Nano Lett* **2016**, *16* (2), 1076-1080.

(25) Kishino, K.; Mizuno, A.; Honda, T.; Yamada, J.; Togashi, R. Improving the luminous efficiency of red nanocolumn μ-LEDs by reducing electrode size to ϕ2.2 μm. *Applied Physics Express* **2023**, *17* (1).

(26) Wang, G. T.; Li, Q.; Wierer, J. J.; Koleske, D. D.; Figiel, J. J. Top–down fabrication and characterization of axial and radial III-nitride nanowire LEDs. *physica status solidi (a)* **2014**, *211* (4), 748-751.

(27) Tseng, W. J.; Gonzalez, M.; Dillemans, L.; Cheng, K.; Jiang, S. J.; Vereecken, P. M.; Borghs, G.; Lieten, R. R. Strain relaxation in GaN nanopillars. *Applied Physics Letters* **2012**, *101* (25).

(28) Zhang, L.; Lee, L.-K.; Teng, C.-H.; Hill, T. A.; Ku, P.-C.; Deng, H. How much better are InGaN/GaN nanodisks than quantum wells—Oscillator strength enhancement and changes in optical properties. *Applied Physics Letters* **2014**, *104* (5).

(29) Muziol, G.; Turski, H.; Siekacz, M.; Szkudlarek, K.; Janicki, L.; Baranowski, M.; Zolud, S.; Kudrawiec, R.; Suski, T.; Skierbiszewski, C. Beyond Quantum Efficiency Limitations Originating from the Piezoelectric Polarization in Light-Emitting Devices. *ACS Photonics* **2019**, *6* (8), 1963-1971.

(30) Turski, H.; Siekacz, M.; Wasilewski, Z. R.; Sawicka, M.; Porowski, S.; Skierbiszewski, C. Nonequivalent atomic step edges—Role of gallium and nitrogen atoms in the growth of InGaN layers. *Journal of Crystal Growth* **2013**, *367*, 115-121.

(31) Skierbiszewski, C.; Turski, H.; Muziol, G.; Siekacz, M.; Sawicka, M.; Cywiński, G.; Wasilewski, Z. R.; Porowski, S. Nitride-based laser diodes grown by plasma-assisted molecular beam epitaxy. *Journal of Physics D: Applied Physics* **2014**, *47* (7).

(32) *SiLENSe 6.5.1* **Accessed on 18th April 2025**, https://str-soft.com/devices/silense/.

(33) Martín, G.; López-Conesa, L.; Pozo, D. d.; Portillo, Q.; Doundoulakis, G.; Georgakilas, A.; Estradé, S.; Peiró, F. Precessed electron diffraction study of defects and strain in GaN nanowires fabricated by top-down etching. *Applied Physics Letters* **2022**, *121* (8).

(34) González-Izquierdo, P.; Rochat, N.; Sakowski, K.; Zoccarato, D.; Charles, M.; Borowik, Ł. GaN/InGaN LED Sidewall Defects Analysis by Cathodoluminescence and Photosensitive Kelvin Probe Force Microscopy. *ACS Photonics* **2024**, *11* (5), 2097-2104.

(35) Wong, M. S.; Hwang, D.; Alhassan, A. I.; Lee, C.; Ley, R.; Nakamura, S.; DenBaars, S. P. High efficiency of III-nitride micro-light-emitting diodes by sidewall passivation using atomic layer deposition. *Opt Express* **2018**, *26* (16), 21324-21331.

(36) Ley, R. T.; Smith, J. M.; Wong, M. S.; Margalith, T.; Nakamura, S.; DenBaars, S. P.; Gordon, M. J. Revealing the importance of light extraction efficiency in InGaN/GaN microLEDs via chemical treatment and dielectric passivation. *Applied Physics Letters* **2020**, *116* (25).

(37) Wong, M. S.; Tak, T.; Ni, A. Y.; Nitta, K.; Gandrothula, S.; Kim, J.; Cha, N.; Mishra, U. K.; Speck, J. S.; DenBaars, S. P. Quantitative analysis of leakage current in III-nitride micro-light-emitting diodes. *Applied Physics Letters* **2025**, *126* (4).

(38) Wierer, J. J.; Tansu, N. III-Nitride Micro-LEDs for Efficient Emissive Displays. *Laser & Photonics Reviews* **2019**, *13* (9).

(39) Han, D.-P.; Kim, H.; Shim, J.-I.; Shin, D.-S.; Kim, K.-S. Influence of carrier overflow on the forward-voltage characteristics of InGaN-based light-emitting diodes. *Applied Physics Letters* **2014**, *105* (19).

(40) Liu, Y.; Zhanghu, M.; Feng, F.; Li, Z.; Zhang, K.; Kwok, H. S.; Liu, Z. Identifying the role of carrier overflow and injection current efficiency in a GaN-based micro-LED efficiency droop model. *Opt Express* **2023**, *31* (11), 17557-17568.

(41) Wong, M. S.; Gee, S.; Tak, T.; Gandrothula, S.; Rebollo, S.; Cha, N.; Speck, J. S.; DenBaars, S. P. Optical analysis of III-nitride micro-light-emitting diodes with different sidewall treatments at low current density operation. *Japanese Journal of Applied Physics* **2024**, *63* (4).

(42) Yapparov, R.; Quevedo, A.; Tak, T.; Nakamura, S.; DenBaars, S. P.; Speck, J. S.; Marcinkevičius, S. Impact of threading dislocations on the V-defect assisted lateral carrier injection and recombination in InGaN quantum well LEDs. *Applied Physics Letters* **2025**, *126* (21).

(43) Huang, H.-C.; Chen, S.-M.; Weisbuch, C.; Speck, J. S.; Wu, Y.-R. The Influence of V-Defects, Leakage, and Random Alloy Fluctuations on the Carrier Transport in Red InGaN MQW LEDs. *arXiv preprint arXiv:2501.19020* **2025**.

(44) Shah, J. M.; Li, Y.-L.; Gessmann, T.; Schubert, E. F. Experimental analysis and theoretical model for anomalously high ideality factors ($n \gg 2.0$) in AlGaN/GaN pn junction diodes. *Journal of applied physics* **2003**, *94* (4), 2627-2630.

(45) Koester, R.; Sager, D.; Quitsch, W. A.; Pfingsten, O.; Poloczek, A.; Blumenthal, S.; Keller, G.; Prost, W.; Bacher, G.; Tegude, F. J. High-speed GaN/GaInN nanowire array light-emitting diode on silicon(111). *Nano Lett* **2015**, *15* (4), 2318-2323.

(46) Lee, J.; Chiu, Y. C.; Bayram, C. Efficiency cliff in scaling InGaN light-emitting diodes down to submicron. *Applied Physics Letters* **2025**, *126* (24).

(47) Cheng-Yin Wang, L.-Y. C., Cheng-Pin Chen, Yun-Wei Cheng, Min-Yung Ke, Min-Yann Hsieh, Han-Ming Wu, Lung-Han Peng, and JianJang Huang. GaN nanorod light emitting diode arrays with a nearly constant electroluminescent peak wavelength. *Optics Express* **2008**, *16*, 10549-10556

(48) Mateusz Hajdel, O. G., Anna Feduniewicz-Żmuda, Marta Sawicka, Carsten Richter, Cedric Corley-Wiciak, Mateusz Słowikowski, Nirmal Anand, Sharif Md. Sadaf, Conny Becht, Ulrich T. Schwarz, Grzegorz Muziol. Elastic strain relaxation in InGaN nanopillars on nanoporous GaN. In *Gallium Nitride Materials and Devices XX*, 2025; SPIE: 2025. DOI: https://doi.org/10.1117/12.3040354.

(49) Chlipała, M.; Turski, H.; Żak, M.; Muziol, G.; Siekacz, M.; Nowakowski-Szkudlarek, K.; Fiuczek, N.; Feduniewicz-Żmuda, A.; Smalc-Koziorowska, J.; Skierbiszewski, C. Bottom tunnel junction-based blue LED with a thin Ge-doped current spreading layer. *Applied Physics Letters* **2022**, *120* (17).

(50) Hwang, D.; Mughal, A.; Pynn, C. D.; Nakamura, S.; DenBaars, S. P. Sustained high external quantum efficiency in ultrasmall blue III–nitride micro-LEDs. *Applied Physics Express* **2017**, *10* (3).

(51) Slawinska, J.; Muziol, G.; Kafar, A.; Skierbiszewski, C. Lateral Carrier Diffusion in Ion-Implanted Ultra-Small Blue III-Nitride MicroLEDs. *ACS Appl Mater Interfaces* **2025**, *17* (4), 6473-6479.

(52) David, A.; Young, N. G.; Lund, C.; Craven, M. D. Review—The Physics of Recombinations in III-Nitride Emitters. *ECS Journal of Solid State Science and Technology* **2019**, *9* (1).

(53) Vogl, F.; Avramescu, A.; Lex, A.; Waag, A.; Hetzl, M.; von Malm, N. Enhanced forward emission by backside mirror design in micron-sized LEDs. *Opt Lett* **2024**, *49* (18), 5095-5098.

(54) Coulon, P. M.; Pugh, J. R.; Athanasiou, M.; Kusch, G.; Le Boulbar, E. D.; Sarua, A.; Smith, R.; Martin, R. W.; Wang, T.; Cryan, M.; et al. Optical properties and resonant cavity modes in axial InGaN/GaN nanotube microcavities. *Optics Express* **2017**, *25* (23).

(55) Oto, T.; Aihara, A.; Motoyama, K.; Ishizawa, S.; Okamoto, K.; Togashi, R.; Kishino, K. Plasmonic red-light-emission enhancement by honeycomb-latticed InGaN/GaN ordered fine nanocolumn arrays. *Applied Physics Express* **2023**, *16* (11).

(56) Oto, T.; Namazuta, M.; Hayakawa, S.; Okamoto, K.; Togashi, R.; Kishino, K. Comparison of surface plasmon polariton characteristics of Ag- and Au-based InGaN/GaN nanocolumn plasmonic crystals. *Applied Physics Express* **2021**, *14* (10).

(57) *Lumerical Inc.* *(https://optics.ansys.com/hc/en-us/articles/1500007184901-Lumerical-Citation-Instruction).*

## FIGURE CAPTIONS

**Figure 1**. (a) Schematic of the complete PAMBE-grown InGaN single quantum well LED heterostructure with InGaN barriers on hydride vapor phase epitaxy (HVPE) grown bulk *n*-GaN substrate. (b) scanning electron micrograph (SEM) image of the top-down fabricated nanopixel array (450 nm diameter, 700 nm pitch) post-ICP-RIE etching. (c) SEM image of the same nanopixel array after KOH-based wet etching for sidewall damage removal. (d) Simulated energy band diagram showing spontaneous and piezoelectric polarization effects at equilibrium (0 V bias) conditions.

**Figure 2**. (a) A 45° high-magnification SEM image of the planarized and etched-back nanopixels passivated with ALD $Al_2O_3$ and SOG layers. (b) a 70° high-magnification SEM image of *p*-Contact deposited nanopixel device, showing Ni/Au, p-$In_{0.15}Ga_{0.85}N$ and ALD deposited $Al_2O_3$ layers. (c) Current density–voltage (J–V) characteristics of the fabricated 500 nm nanopixel LED array. Inset illustrates the J-V plot in log scale showing ultra-low leakage currents at reverse bias (~0.01 A/cm² at –10 V). (d) Diode ideality factor versus the forward voltage of the parallelly addressed InGaN nanopixel device.

**Figure 3**. (a) Monte Carlo simulation of the predicted percentage of dislocation-free pixels as a function of pixel diameter, assuming an initial dislocation density of $5\times10^{6}$ $cm^{-2}$ corresponding to HVPE-grown single-crystal GaN substrates. Dashed lines represent diameters of ~500 nm, and 20 $\mu$m pixels. (b) Predicted dislocation-free pixel percentages for varying initial dislocation densities ($5\times10^{6}$ to $1\times10^{9}$ $cm^{-2}$) as a function of pixel diameter, to highlight the impact of top-down nanostructuring across substrates and epi-layer qualities. (c) Schematic illustration showing how top-down nanostructuring effectively isolates threading dislocations, resulting in reduced and/or virtually dislocation-free nanopixels depending on the initial planar epi-layer quality.

**Figure 4.** (a) Electroluminescence (EL) spectra of the nanopixel LED at varying injection current densities. (b) Peak emission wavelength as a function of current density, showing a ~14 nm QCSE induced blue-shift

and full-width half maximum (FWHM) of the EL spectra across injection levels, showing spectral narrowing at high injection.

**Figure 5**. (a) Poisson–Schrödinger simulated energy band diagrams of the $In_{0.179}Ga_{0.821}N$ QW for different degrees of relaxation (DOR). (b) Electron and hole wavefunction distributions for each DOR, showing improved overlap with increasing strain relaxation. (c) Left axis: Calculated $< e1h1 >$ wavefunction overlap versus DOR, highlighting QCSE suppression trends for various QW degrees of relaxation. Right axis: Calculated EL peak wavelength versus DOR.

**Figure 6.** (a) Measured optical output power density of the nanopixel $\mu$LED array compared to planar $\mu$LEDs. Inset shows the microscope image of the device operating at ~100 A/cm$^2$. (b) On-wafer external quantum efficiency (EQE) vs. injection current density, fitted with the ABC model. Inset: fitted ABC model parameters and estimated light extraction efficiency (LEE ≈ 42%).

**Figure 7.** FDTD-simulated light extraction efficiency (LEE) of InGaN nanopixel arrays vs. planar structures over 445–455 nm. Nanopixels (~450 nm diameter, 1 $\mu$m height) in a 700 nm pitch square lattice on GaN or sapphire substrates show ~40% higher extraction efficiency, independent of substrate material.

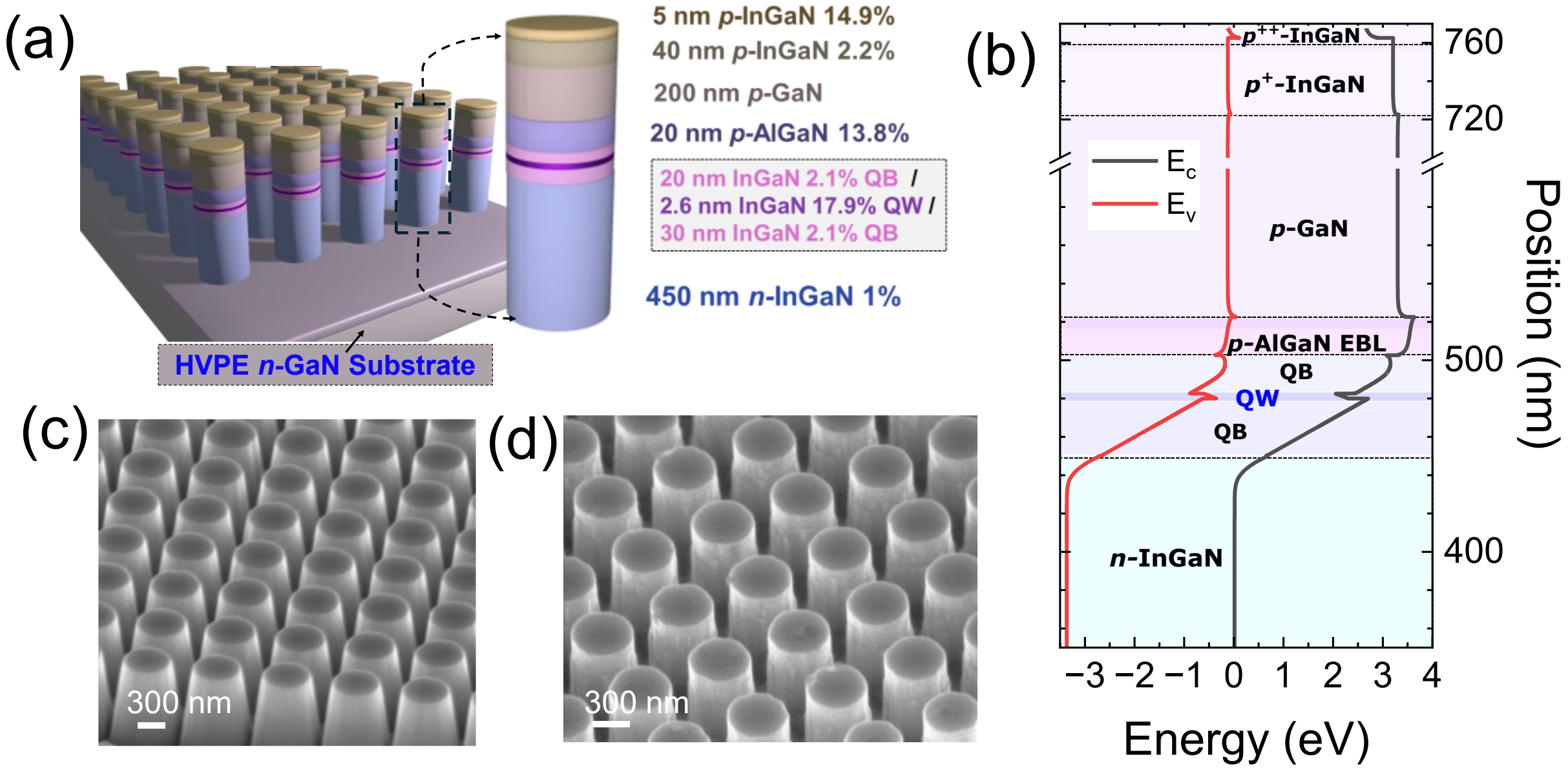
(a)
5 nm p-InGaN 14.9%
40 nm p-InGaN 2.2%
200 nm p-GaN
20 nm p-AlGaN 13.8%
20 nm InGaN 2.1% QB /
2.6 nm InGaN 17.9% QW /
30 nm InGaN 2.1% QB
450 nm n-InGaN 1%
HVPE n-GaN Substrate
(b)
p++-InGaN
p+-InGaN
p-GaN
p-AlGaN EBL
QB
QW
QB
n-InGaN
Ec
Ev
760
720
500
400
Position (nm)
−3 −2 −1 0 1 2 3 4
Energy (eV)
(c)
300 nm
(d)
300 nm

**Figure 1**

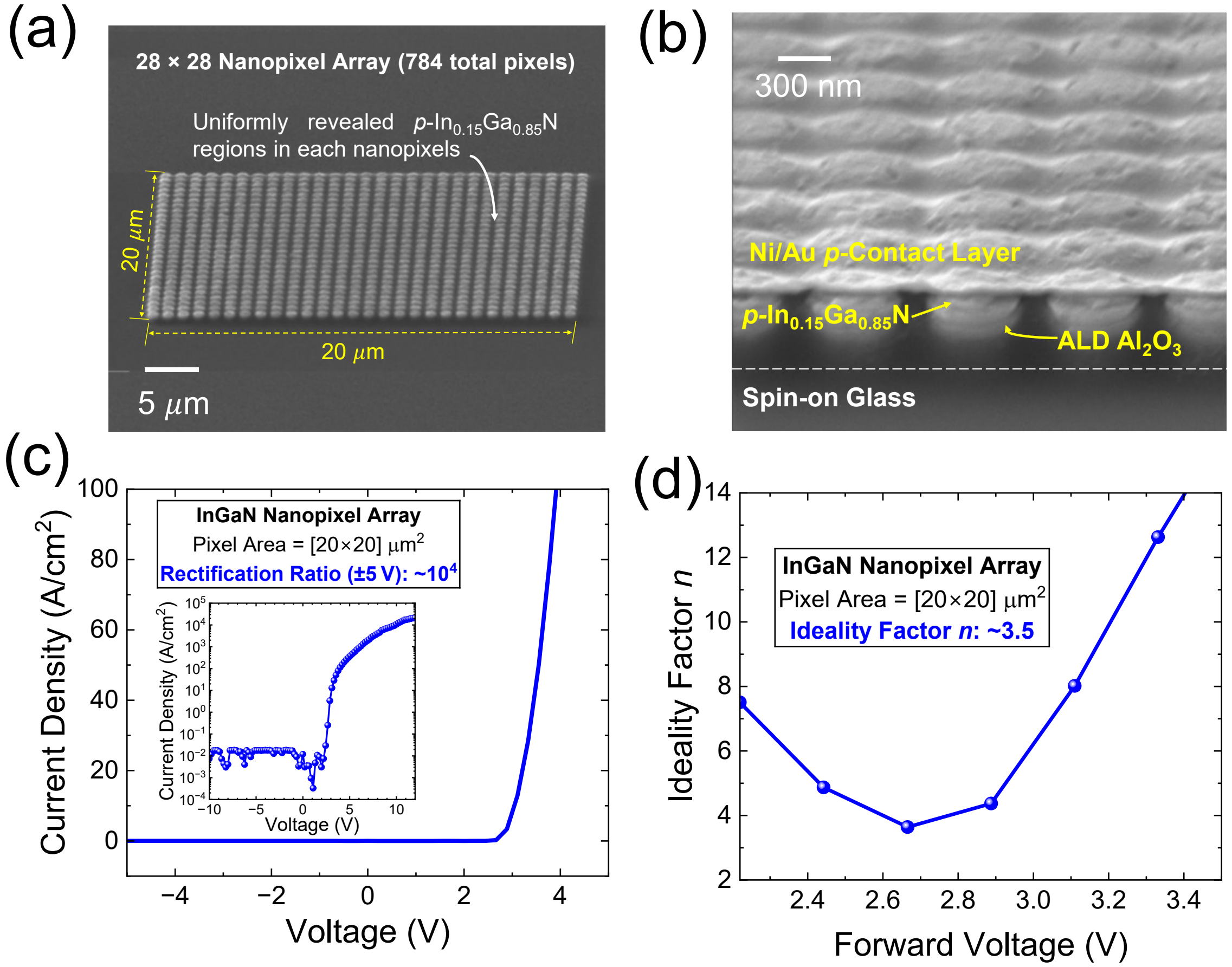
(a)
28 × 28 Nanopixel Array (784 total pixels)
Uniformly revealed p-In0.15Ga0.85N regions in each nanopixels
20 μm
20 μm
5 μm
(b)
300 nm
Ni/Au p-Contact Layer
p-In0.15Ga0.85N
ALD Al2O3
Spin-on Glass
(c)
InGaN Nanopixel Array
Pixel Area = [20×20] μm2
Rectification Ratio (±5 V): ~10^4
Current Density (A/cm2)
Voltage (V)
(d)
InGaN Nanopixel Array
Pixel Area = [20×20] μm2
Ideality Factor n: ~3.5
Ideality Factor n
Forward Voltage (V)

**Figure 2**

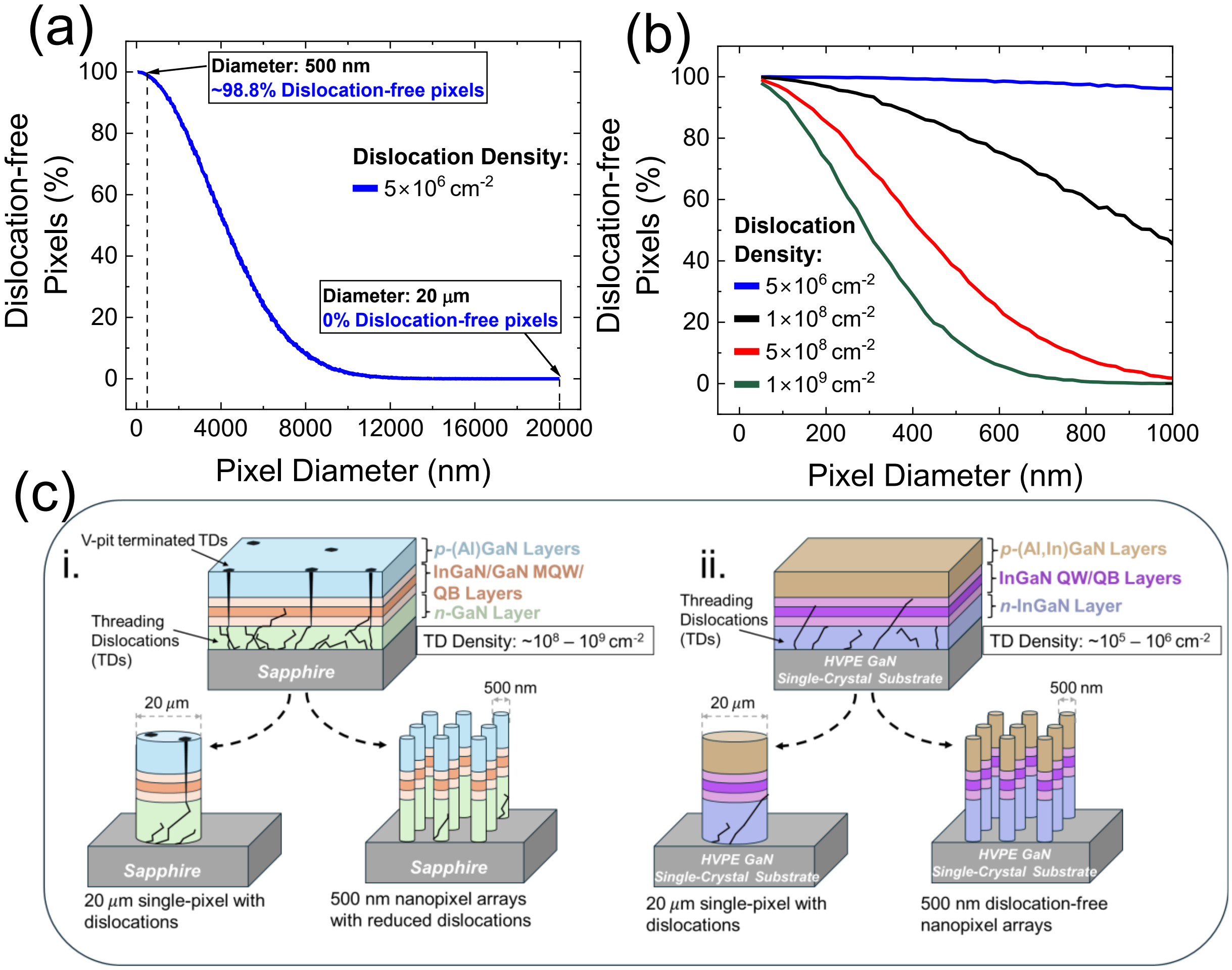

**Figure 3**

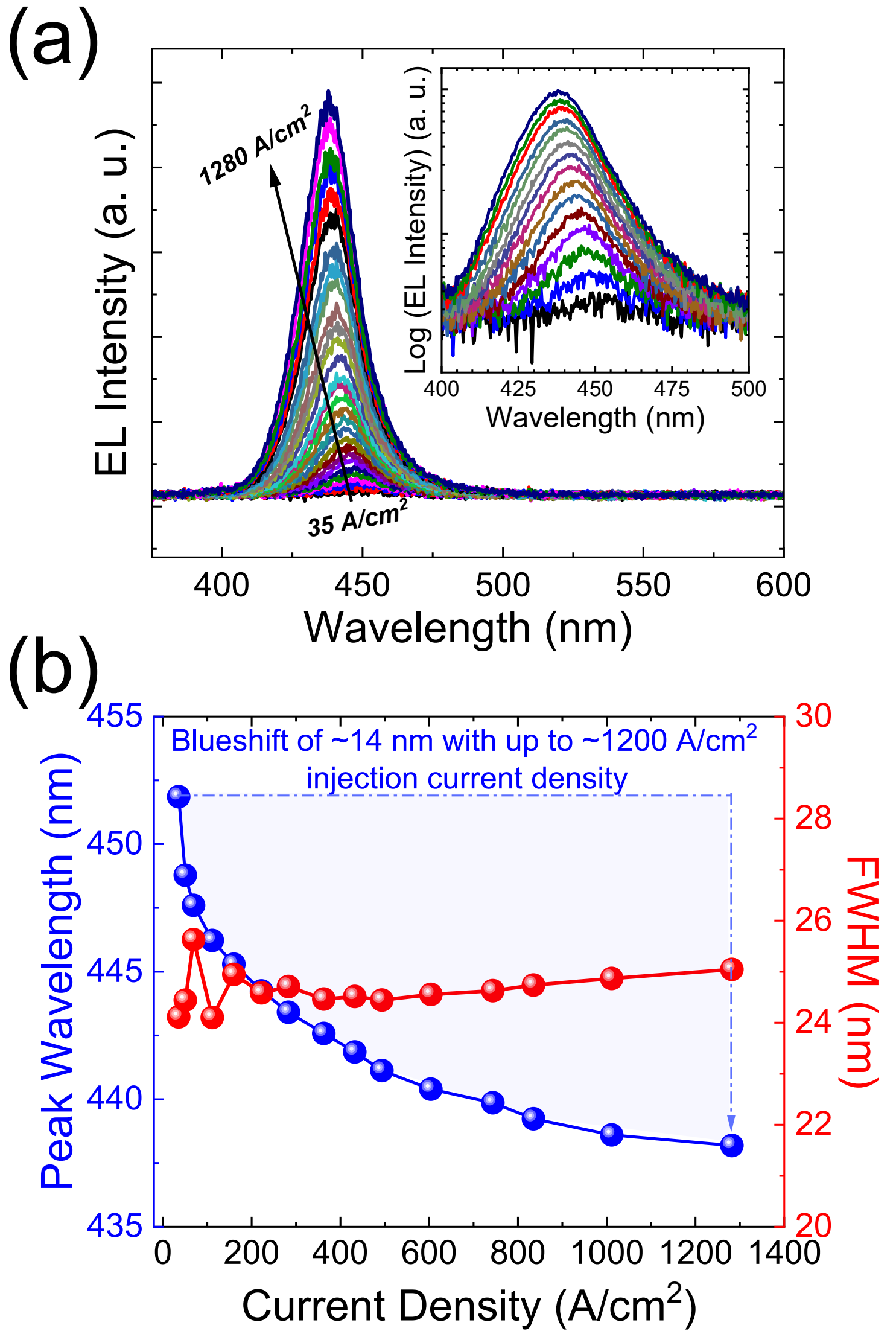
(a)
1280 A/cm2
35 A/cm2
EL Intensity (a. u.)
Wavelength (nm)
400
450
500
550
600
Log (EL Intensity) (a. u.)
400
425
450
475
500
Wavelength (nm)
(b)
Blueshift of ~14 nm with up to ~1200 A/cm2
injection current density
Peak Wavelength (nm)
FWHM (nm)
Current Density (A/cm2)
455
450
445
440
435
30
28
26
24
22
20
0
200
400
600
800
1000
1200
1400

**Figure 4**

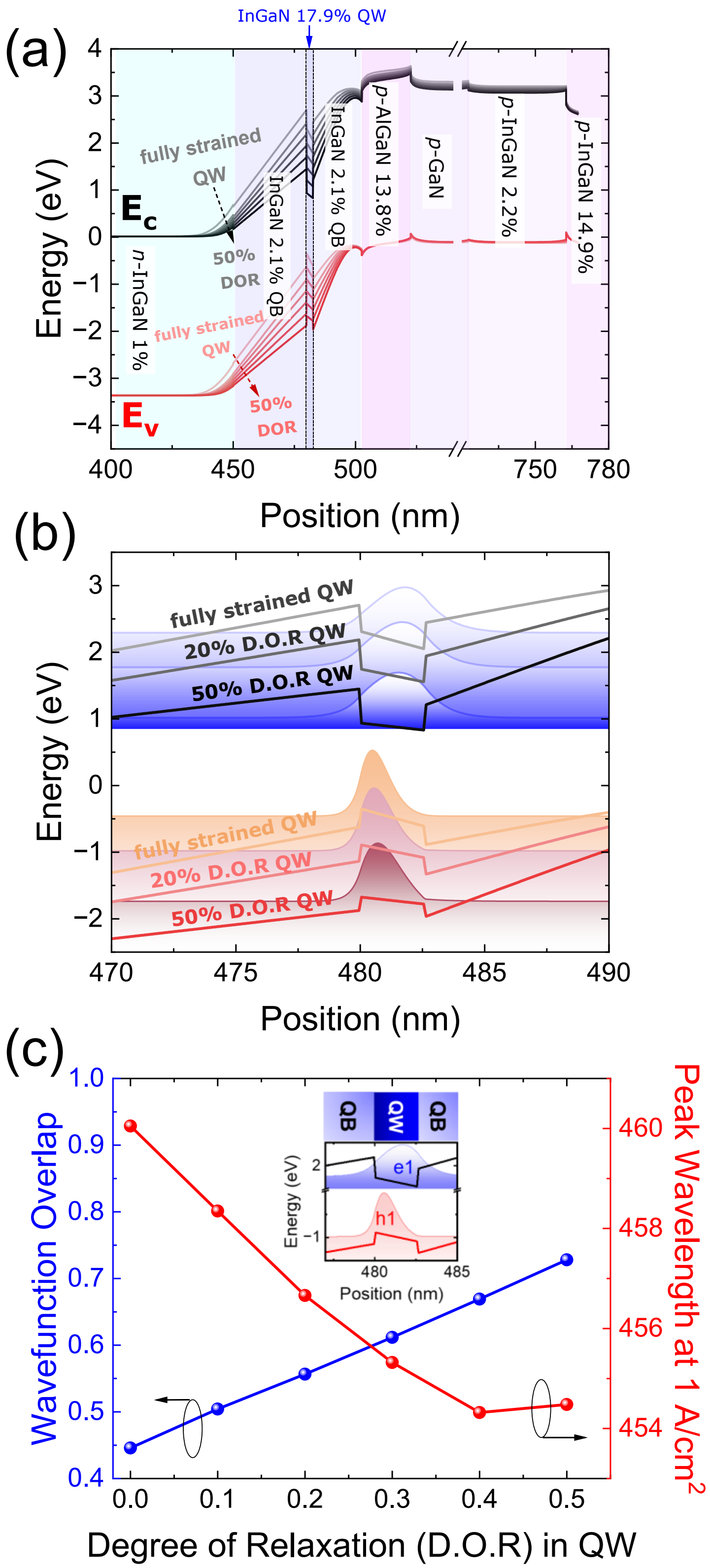

**Figure 5**

| ***Degree of Relaxation (DOR) in QW*** | ***EL Peak WL @~1 A/cm$^2$*** | ***EL Peak WL @~1000 A/cm$^2$*** | ***Peak WL shift (Δλ)*** |
|---|---|---|---|
| 0% (fully strained) | 460.05 nm | 442.80 nm | 17.25 nm |
| 10% | 458.35 nm | 443.27 nm | 15.08 nm |
| 20% | 456.66 nm | 443.11 nm | 13.55 nm |
| 30% | 455.32 nm | 442.95 nm | 12.40 nm |
| 40% | 454.32 nm | 442.95 nm | 11.37 nm |
| 50% | 454.48 nm | 443.43 nm | 11.05 nm |

**Table 1.**

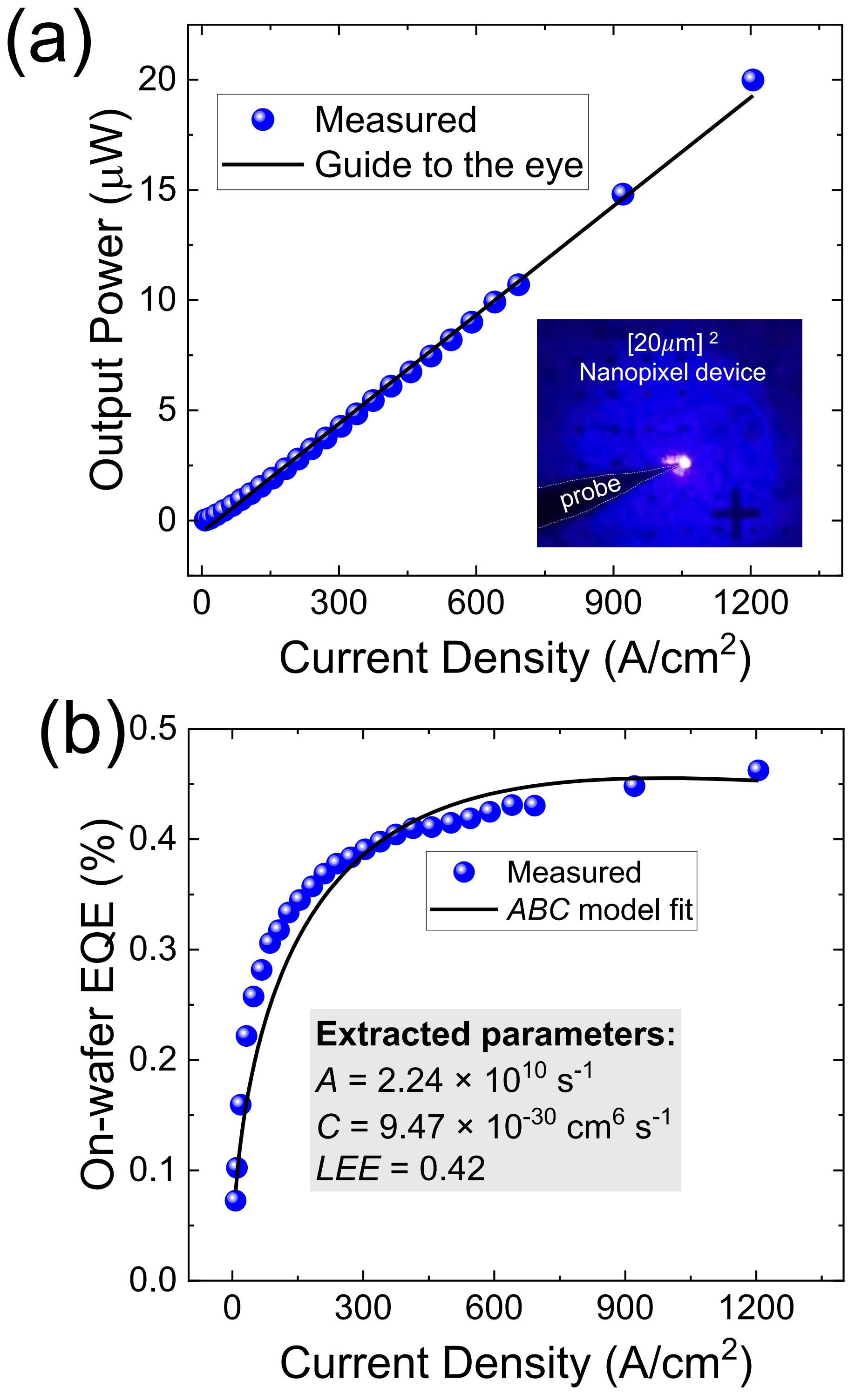

**Figure 6**

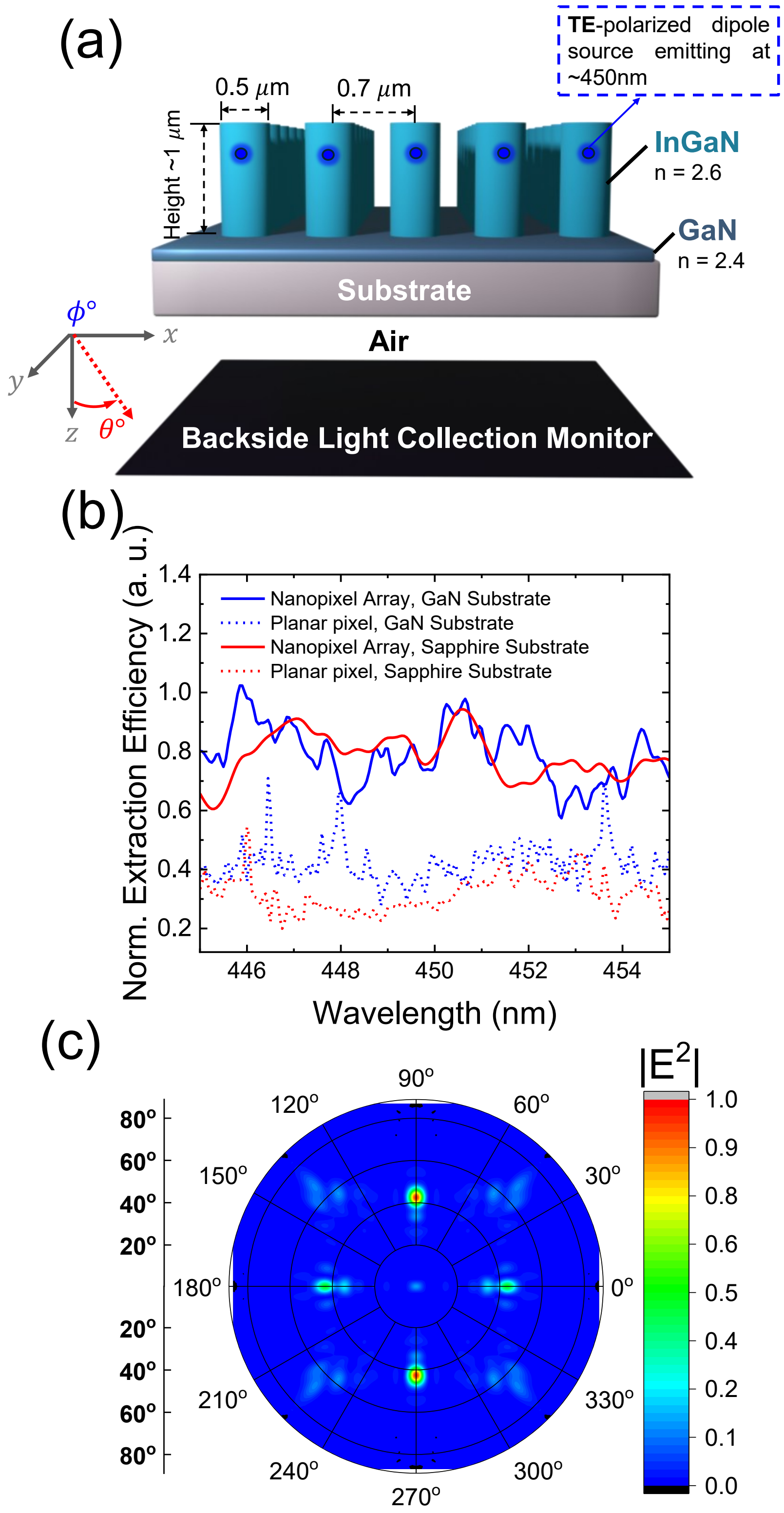
(a)
TE-polarized dipole source emitting at ~450nm
0.5 μm
0.7 μm
Height ~1 μm
InGaN
n = 2.6
GaN
n = 2.4
Substrate
Air
Backside Light Collection Monitor
ϕ°
x
y
z
θ°
(b)
Norm. Extraction Efficiency (a. u.)
Nanopixel Array, GaN Substrate
Planar pixel, GaN Substrate
Nanopixel Array, Sapphire Substrate
Planar pixel, Sapphire Substrate
Wavelength (nm)
(c)
|E2|
90°
120°
60°
150°
30°
180°
0°
210°
330°
240°
300°
270°
80°
60°
40°
20°

**Figure 7**